\documentclass[twocolumn,showpacs,preprintnumbers,amsmath,amssymb]{revtex4}
\usepackage{graphicx}
\usepackage{dcolumn}
\usepackage{bm}
\begin{document}
\title{Understanding H-defect complexes in ZnO}
\author{R. Vidya$^{1,2}$}
\email{vidya.ravindran@smn.uio.no}
\author{P. Ravindran$^{1,3}$}
\author{H. Fjellv{\aa}g$^{1}$}
\address{$^{1}$Center for Materials Science and Nanotechnology and Department of Chemistry, University of Oslo, Box 1033
Blindern, N-0315 Oslo, Norway }
\address{$^{2}$Institute of Mathematical Sciences, CIT Campus, Taramani, Chennai 600 113, India}
\address{$^{3}$Department of Physics, Central University of Tamil Nadu, Thiruvarur, Tamil Nadu, 610 001, India }
\date {\today}
\begin{abstract}
{\sloppy  From state-of-the-art density-functional calculations
using hybrid functionals we show that, persistent $n$-type conductivity in ZnO is due to
defect complexes formed between H with intrinsic and extrinsic
defects. H exhibits cationic, anionic, and electrically-inactive character on
interacting with defects in ZnO. The electrically-inactive molecular
hydrogen can contribute to $n$-type conductivity in ZnO by
activating deep donor levels into shallow levels. By calculating
local vibrational mode frequencies, we have identified origins of many H-related
Raman and infra-red frequencies and thus confirmed the amphoteric behavior
of H.}

\end{abstract}

\maketitle ZnO is one of the the most studied wide band-gap
(E$_{g}$) semiconductors because of its wide range of
applications. Though as-prepared samples of ZnO contain many
impurities, interaction of ubiquitous impurity like H with ZnO has
many important technological implications. For example, H is
expected to become an environmentally-benign fuel in future and
nanophase ZnO can be a potential sensor for H because of its
excellent characteristics~\cite{chen05}. Moreover, ZnO is considered as a
prospective material for hydrogen storage in future~\cite{eizek08}
because high amount of hydrogen (30.5 at.\%) can be introduced
into ZnO crystals by electrochemical charging. The stability of ZnO
against hydrogen plasma makes it interesting also for solar cell
applications where ZnO is used as a transparent-conducting
electrode~\cite{beyer07}. Therefore it is
vital to understand the influence of H on properties of ZnO.
\par
ZnO exhibits $n$-type conductivity even without deliberate doping.
Introduction of hydrogen into ZnO at elevated temperatures was
shown to give rise to the $n$-type
conductivity~\cite{mollwo54,thomas56}. First-principles
investigations~\cite{walle00,janotti07} substantiated that
interstitial H (H$_{i}$) and substitutional H (H$_{O}$) are indeed
shallow donors and contribute to the $n$-type conductivity. By
annealing hydrogenated ZnO samples, 78\% of the free carriers were
eliminated near 150$^{o}$C and the remainder between
500$-$700$^{o}$C~\cite{shi04}. However H$_{i}$ is
shown~\cite{wardle06,ip03} to be unstable above room temperature,
making it difficult to understand how hydrogen alone can give rise
to conductivity at high temperatures. Moreover, the donor levels
from H$_{i}$ and H$_{O}$ are shown to be resonant inside
conduction band whereas various experimental
techniques~\cite{thomas56,ip03,hutson57,hofmann02,look03,kassier08,lavrov09,weber10},
observed a H-related shallow donor level at 25$-$50 meV below
conduction band minimum (CBM) whose origin is not clearly
understood.
\par
In addition, many vibrational frequencies have been measured in
ZnO by infra-red (IR) spectroscopy whose origins are still
puzzling. For example, the prominent peak at 3326 was earlier
believed to be related to H$_{i}$ at anti-bonding
(AB$_{O,\perp}$)~\cite{shi04} and later assigned to
V$_{Zn}$+H$_{i}$~\cite{herklotz10} and Ca$_{Zn}$+H$_{i}$
complex~\cite{li08}. The peak at 3611 cm$^{-1}$ is assigned to
H$_{i}$ at bond-centered (BC$_{\parallel}$) site ~\cite{lavrov09},
but impurity-hydrogen complex was not ruled out as this peak was
stable up to 350$^o$C in annealing experiments~\cite{lavrov02}.
IR and Raman spectroscopy can provide valuable clues about the nature of
a defect, from frequency and symmetry of the observed modes. However, comparison
of these values with first-principles calculations can provide a conclusive
microscopic identification of a defect~\cite{walle06}.
\par
We have performed first-principles calculations using the
projected augmented plane-wave method~\cite{blochl94} implemented
in the Vienna {\it ab initio} simulation package (VASP)~\cite{vasp}. Complete
structural optimizations with H  at various
interstitial/substitutional positions, without and with impurities
like Li, B, C, N, Al, and Ga in different charge states were done.
Supercells of 192 atoms with a plane-wave energy cutoff of 550\,eV
were used. The atomic geometry was optimized by force as well as
stress minimizations with convergence criteria of 10$^{-6}$\,eV
per unit cell for total energy and $\le$ 1
meV\,\rm{\,{\AA}}$^{-1}$ for forces. Exchange and correlation
effects are treated under the
generalized-gradient-approximation~\cite{perdew96} with
Perdew-Burke-Ehrenkof functional. To identify the origin of
experimentally-observed IR and Raman-active phonon mode
frequencies, we have derived the local vibrational mode (LVM)
frequencies from ``frozen-phonon" calculations. The concerned
atoms are displaced in steps of 0.001 {\AA} to a maximum of 0.005
{\AA} in both directions to account for possible anharmonic
effects. More details are given in Ref.~\onlinecite{ravi03} where
the calculated frequencies  are shown to be within 10 cm$^{-1}$
from experimental values.
\par
Finally one set of structure optimization and electronic structure
calculation was carried out using the Heyd-Scuseria-Ernzerhof
(HSE) hybrid functional~\cite{heyd03} with a screening parameter
of $a$ = 0.375 which reproduced the experimental structural parameters
and band-gap value for ZnO.
The total energy obtained using HSE functional is used to derive
defect formation energy and thermodynamic transition levels of
various defects as described in Refs.~\onlinecite{vidya11,oba08}.

\begin{figure}
\includegraphics[scale=0.575]{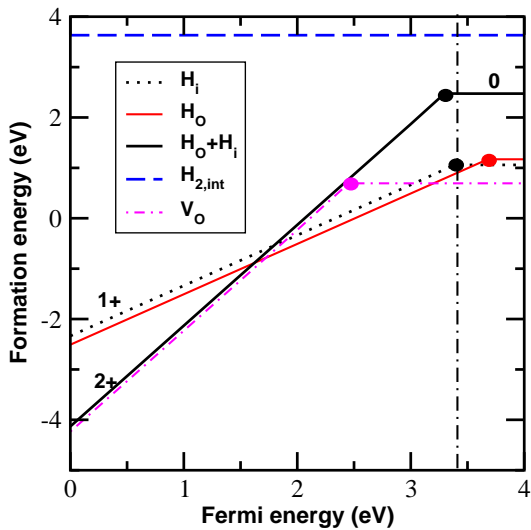}
\caption{\label{fig:trans_hs} (Color online) Formation energy of
defects obtained using hybrid functionals (under  Zn-rich
conditions). The dots indicate charge-transition points and the
charge states are given on the illustration.}
\end{figure}
Hydrogen is shown to be commonly present in as-grown ZnO samples.
Diffusion of H$_{i}$ due to annealing of sample at 150$^o$C
~\cite{shi04} was suggested to form interstitial H$_2$ molecule
which is "hidden" from experimental techniques like IR, ESR
measurements etc. The Raman frequency for this so-called "hidden"
hydrogen is recently measured to be 4145
cm$^{-1}$~\cite{lavrov09}. In order to understand formation
of "hidden" hydrogen better, we modeled H at six different interstitial positions
(H$_{i}$), H at oxygen vacancy (H$_{O}$), and H$_{2}$ molecule in the interstitial channel
(H$_{2,int}$) oriented along $x, y, z$ directions. We found that H$_{2,int}$ along $z$-direction in neutral state has
the lowest formation energy among its counterparts oriented along $x$ and $y$
directions.
However, the formation energy of H$_{i}$ and H$_{O}$ is lower compared to that
of H$_{2,int}$.  The calculated Raman frequency
for H$_{2,int}$ along $z$-direction is 4047 cm$^{-1}$, in agreement with
previous theoretical study~\cite{wardle06}. It can be seen from
Fig.~\ref{fig:trans_hs} that H$_{2,int}$
has the highest formation energy among other H-related defects
considered.
\par
\begin{figure*}
\includegraphics[scale=0.5]{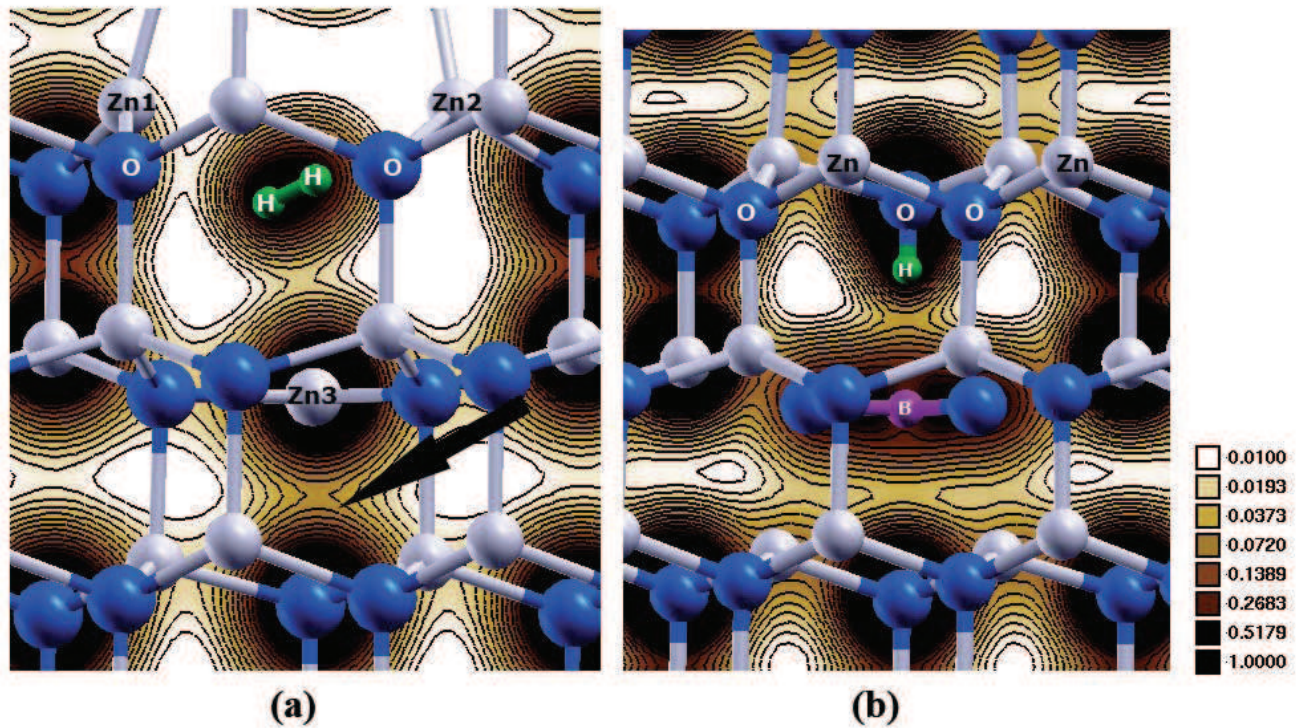}
\caption{\label{fig:crys} (Color online) Electronic charge density
of ZnO showing (a) electrically neutral H$_O$+H$_i$; Bonding
states arising due to Zn displacement are shown by arrow  (b)
cationic H in B$_{Zn}$+H$_{i}$ at BC$_{\parallel}$.}
\end{figure*}

It has been well established that the oxygen vacancy (V$_{O}$) is
the dominant intrinsic defect in ZnO~\cite{oba08, walle00}. When
V$_{O}$ is formed, electrons from dangling bonds of
surrounding Zn sites are localized at the vacancy~\cite{vidya11}.
We have shown~\cite{vaj05} that H prefers to occupy sites where
non-bonding localized electron density is maximum and where it
could obtain maximum screening~\cite{janotti07}. In conjunction,
when H occupies the V$_{O}$ (H$_O$), the surrounding Zn atoms are
relaxed towards H$_{O}$. As V$_{O}$ prefers to be in 2+ state, we
introduced another interstitial H close to H$_O$, and thus
modelled two configurations: (i) H$_O$ and a H$_{i}$ at
BC$_{\parallel}$ and (ii) H$_O$ and a H$_{i}$ at AB$_{O,\perp}$
(see Fig.~\ref{fig:crys}c). Interestingly, the H$_{i}$ placed at a
distance of 0.98 {\AA} from H$_O$, relaxes towards H$_O$ and forms
a complex with a H$_O$$-$H$_i$ distance of 0.76 {\AA}. It is worthwhile to
note that the H$-$H distance in gaseous H$_2$ is 0.74 {\AA}. Hence the H$_O$+H$_{i}$
complex can be visualized as a H$_{2}$ molecule trapped at a V$_{O}$. Formation
energy of this complex is nearly 1.25 eV lower than the isolated H$_{2,int}$ molecule
(Fig.~\ref{fig:trans_hs}).
\begin{figure}
\includegraphics[scale=0.7]{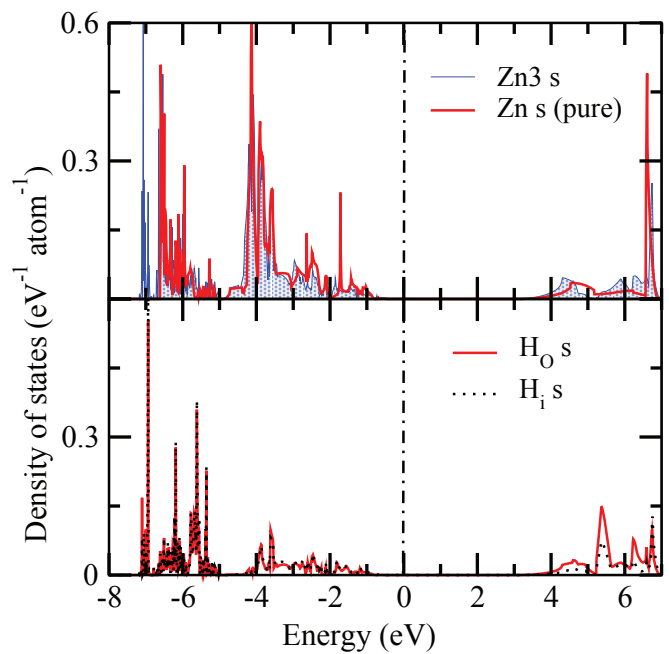}
\caption{\label{fig:DOS} (Color online) Density of state for $s$-orbitals of H
in the H$_O$+H$_{i}$ complex and $s$-orbitals of the neighboring Zn atoms that is
displaced along $z$-axis (see text). For comparison, Zn-$s$ orbitals in pure
ZnO is shown in red (solid) lines.}
\end{figure}
\par
Interestingly, the calculated Raman frequencies for
H$_O$+H$_{i}$(BC$_{\parallel}$) and H$_O$+H$_{i}$(AB$_{O,\perp}$)
are 4103 and 4108 cm$^{-1}$, respectively, in better agreement
with experimental value than that for H$_{2,int}$ (see Table 1).
The calculated Raman frequency for D$_O$+D$_{i}$ complex (2908
cm$^{-1}$) is also in good agreement with experimental value of
2985 cm$^{-1}$. Thus, H$_O$+H$_{i}$ complex could be a better
candidate for the so-called "hidden" hydrogen in ZnO, in line
with the recent calculations~\cite{du11}.
\par
In addition to the 4145 cm$^{-1}$ line, a weaker component
blue-shifted by 8 cm$^{-1}$ was also observed~\cite{lavrov09},
believed to be due to the ortho-para splitting of the interstitial
H$_{2}$ molecule in ZnO. The calculated formation energy of
H$_O$+H$_{i}$(BC$_{\parallel}$) is only slightly higher (2 meV) than
H$_O$+H$_{i}$(AB$_{O,\perp}$). Therefore H$_{i}$ in the
H$_O$+H$_{i}$ complex can easily move from the AB$_{O,\perp}$ site
to the BC$_{\parallel}$ site, leading to a difference in Raman
frequency of 5 cm$^{-1}$. Hence, in addition to the ortho-para
splitting of the H$_{2}$ molecule (trapped at V$_{O}$), different
geometric orientations of the H$_O$+H$_{i}$ complex could also
lead to the observed blue-shift of the Raman frequency.
\par
The H$_O$+H$_{i}$ complex has the lowest formation energy (up to
mid-gap value; Fig.~\ref{fig:trans_hs}) among H-related defects
considered. While the 1+ to 0 transition of H$_{i}$ occurs exactly
at CBM, the H$_{O}$ has the same at 0.3 eV above the CBM. On the
other hand, the H$_O$+H$_{i}$ complex is stable in 2+ state for
most of the Fermi energy values and the 2+ to 0 transition occurs at 50
meV below CBM (shallow donor level). This could explain the electrical conductivity
measurements (40$-$51 meV) made in 1950s~\cite{thomas56,hutson57}
and the following experimental observations: The ionization energy
of H in ZnO is 35$\pm$5 meV by EPR
measurements~\cite{ip03,hofmann02}. The Hall measurements on
vapor-phase grown samples showed a H donor level at 30-40 meV
which increased to 50 meV upon increasing temperature.
Hydrogen-implanted hydrothermal-grown ZnO films and ZnO annealed
in Zn-rich (O-deficient) conditions also showed a donor level
$\approx$ 45 meV that was stable at high
temperatures~\cite{kassier08,weber10}. The photoluminescence
spectra exhibit a sharp excitonic feature at 3.363 eV (known as
$I_4$ line; optical fingerprint of H in vapor-phase grown
ZnO~\cite{look03}) very close to where we observe the 2+ to 0
transition of the H$_O$+H$_{i}$ complex.

\par
Usually H$_2$ is believed to be electrically inactive, whereas the
present study shows that a shallow donor level is created when
H$_2$ occupies V$_{O}$. To understand the mechanism leading to the
formation of this shallow level we analyzed the electronic
structure in detail. The calculated DOS using HSE functional
(E$_{g}$ = 3.3 eV) show that $s$-orbitals of H$_O$ and H$_{i}$ are
well-localized with sharp peaks between $-$8 to $-$6 eV, implying
that the H$_O$+H$_{i}$ complex is electrically-inactive. The
charge density surrounding the H$_O$+H$_{i}$ complex clearly
substantiates its molecular-like character (Fig.~\ref{fig:crys}a).
\par
To obtain more insight, the DOS of Zn atoms surrounding the H$_O$+H$_{i}$
complex is compared with that of pure ZnO (shaded DOS in Fig.~\ref{fig:DOS}).
The Zn-$p$ and Zn-$d$ states of Zn atoms surrounding the complex
are not changed very much compared to those in pure ZnO. However,
the unoccupied Zn-$s$ levels are shifted to lower energy from CBM.
When H$_2$ occupies the V$_{O}$, the surrounding Zn atoms are
relaxed outwards from V$_{O}$. The bond-lengths of Zn1, Zn2, and
Zn4 with their neighbors increase by nearly 24.5\% compared to
that in pure ZnO. Zn3 moves away along $z$-axis by 28.3\% to
compensate the repulsive interaction from electrons at the
H$_O$+H$_{i}$ complex. Therefore Zn3 occupies an interstitial site
and becomes co-planar with the neighboring O atoms. The charge density
clearly shows an overlap interaction of Zn-$s$ states with O-$s$ states along
$z$ direction (see arrow in
Fig.~\ref{fig:crys}a) as a result of the displacement of the Zn3 atom.
In agreement with the above inference, the
muon-electron contact hyperfine interaction showed~\cite{cox01} a
shallow level at 60 $\pm$ 10 meV below CBM with a highly dilated
electron wave function of Zn-4$s$ character. Thus the
electrically-inactive H$_2$ molecule at V$_{O}$ indirectly
introduces an overlap interaction between neighboring Zn and O,
which activates the deep donor level from V$_{O}^{2+}$ into a
shallow level (Fig.~\ref{fig:trans_hs}).
\par
Our nudged-elastic band (NEB) calculations show that the
H$_O$+H$_{i}$ complex is 0.38 eV higher in energy than isolated
H$_O$ and H$_i$. Moreover, the reaction barrier to form the
H$_O$+H$_{i}$ complex from isolated H$_O$ and H$_i$ is 0.92 eV
indicating that it can be formed at high temperatures. The
migration barrier for a H$_{2,int}$ to occupy a V$_{O}$ is
calculated to be less than 19 meV, indicating that molecular H
prefers to occupy the V$_{O}$ than to be at the interstitial
channel. As V$_{O}$ is formed at elevated temperature, H$_2$
molecule can occupy the V$_{O}$ to form the H$_O$+H$_{i}$ complex,
consistent with experimental observation~\cite{shi04}. It may be
noted that the experimentally established~\cite{thomas56,mollwo54}
activation energy of a H-related donor at 40 meV observed at high
temperature is 0.91$-$1.12 eV, in good agreement with our finding.
This implies that the H$_O$+H$_{i}$ complex could be a meta-stable
species and could explain the conductivity measurements on ZnO at
high temperatures.

\begin{table}
\caption{Local vibrational mode frequencies (in cm$^{-1}$) of
hydrogen in ZnO. The identified origin of the frequencies, the
corresponding H positions and orientations are given. }
\begin{ruledtabular}
\begin{tabular}{lccccc}
\multicolumn{2}{c} {Frequency} &  Origin  & H position & Orientation  \\
Expt. [Ref.]  & Present \\
\hline
4145 [\onlinecite{lavrov09}]   & 4108      & H$_{O}$+H$_{i}$   & AB$_{O,\perp}$ &  -- \\
  -     & 4047     & H$_{2,int}$           & ${\parallel}$c & --\\
  -     & 338      & & H$_{O}$+H$_{i}$     & BC$_{\parallel}$ & -- \\
  -     & 560      & Li$_{Zn}$+H$_{O}$     &  $V_{O}$             & ${\parallel}$c\\
3611  [\onlinecite{lavrov02}]  & 3621     & B$_{Zn}$+H$_{i}$  & BC$_{\parallel}$ & ${\parallel}$c \\
3326 [\onlinecite{mccluskey02}] & 3337   & Al$_{Zn}$+H$_{i}$ & AB$_{O,\perp}$ & ${\perp}$c \\
3349 [\onlinecite{lavrov02}] & 3339     & V$_{Zn}$+2H$_{i}$ & AB$_{O,\perp}$   & ${\perp}$c \\
- & 3352     & Ga$_{Zn}$+H$_{i}$ & BC$_{\parallel}$ & ${\parallel}$c \\
3312 [\onlinecite{lavrov02}]  & 3312     & Al$_{Zn}$+H$_{i}$ &
BC$_{\parallel}$ & ${\parallel}$c \label{table:vibfreq1}
\end{tabular}
\end{ruledtabular}
\end{table}
\par
Many H-related vibrational frequencies have been measured in ZnO by
IR spectroscopy, in particular absorption peaks at 3611 and 3326 cm$^{-1}$
are the most prominent frequencies. As ZnO can be prepared by many
different synthetic techniques and conditions, incorporation of different
impurities becomes unavoidable. Although frequency and symmetry of the
observed modes from IR and Raman spectroscopy can provide valuable clues
about the nature of a defect, comparisons with first-principles calculations
are needed to provide a conclusive microscopic identification of a defect~\cite{walle06}.
In order to isolate the role of individual defects, we have calculated
energetics and IR frequencies of H in different interstitial positions
as well as with company of other impurities.
\par
The calculated LVM of H at various interstitial/substitutional
positions together with other defects/impurities are given in
Table~\ref{table:vibfreq1}. The calculated LVM for isolated
H$_{i}$ at BC$_{\parallel}$, AB$_{\parallel}$, BC$_{\perp}$, and
AB$_{\perp}$ are 3565, 3440, 2540, and 3155 cm$^{-1}$,
respectively. On the other hand, the
calculated LVM for H$_{i}$ at BC$_{\parallel}$ adjacent to B$_{Zn}$
(Boron at V$_{Zn}$; Fig.~\ref{fig:crys}b) gives a
value of 3621 cm$^{-1}$, in excellent agreement with experiment.
Interestingly, the samples grown by CVD have Boron as one of the dominant
impurities which lead to the peak at 3611 cm$^{-1}$~\cite{mccluskey07}.
Moreover, for H$_{i}$ at AB$_{\perp}$ adjacent to Al$_{Zn}$ the calculated
frequency is 3337 cm$^{-1}$, in good agreement with experimental
value. It is noteworthy that SIMS measurements~\cite{mccluskey07}
showed significant concentrations of Al in the melt-grown samples
that have the dominant peak at 3326 cm$^{-1}$~~\cite{jokela03}.
The experimental frequencies are dependent on sample preparation and
annealing conditions~\cite{walle06,jokela07}, substantiating our findings that the observed
hydrogen modes could also arise from impurity-hydrogen complexes.
Interestingly, these complexes are stable in 2+ charge state, implying
their contribution to the $n$-type conductivity in ZnO where H has a cationic character.
Even though we have considered as many impurity+H complexes as possible
within computational limitations, other defect complexes leading to the observed frequencies can not be completely ruled out.
\par
As seen above, H by attaching to O exhibits cationic character with impurities like
B, and Al  (Fig.~\ref{fig:crys}b) like in the
case of proton-conducting oxides. Additional calculations show that anionic H (H$_{O}$ close to Li$_{Zn}$) is energetically more favorable than cationic H in the company of Li$_{Zn}$. When H occupies an intrinsic
defect like V$_{O}$, it takes up anionic
character (as in ionic hydrides) and brings in $n$-type
conductivity at ambient conditions. In contrast, in molecular H
and organic solids, H is neutral and forms strong covalent
bonding. As shown in Fig.~\ref{fig:trans_hs} and
Fig.~\ref{fig:crys}a, this neutral H$_2$ at V$_O$ induces shallow
donor level at high temperatures in ZnO. The frequencies 3611 and
4145  cm$^{-1}$ are measured in the same sample under different
annealing conditions~\cite{lavrov09, lavrov09_1}. Our calculations
show that formation of cationic H (H$_{i}$
close to B$_{Zn}$ and Al$_{Zn}$), neutral (H$_{2}$ at V$_{O}$) and,
anionic H (H$_{O}$ close to Li$_{Zn}$) is
energetically favorable and lead to the often-observed IR and Raman frequencies
in ZnO. This implies that H can exhibit amphoteric behavior in
the same sample under different conditions. This
chemically-adaptable versatile nature of H makes it stable with
the company of both donor and acceptor-type impurities. However, H
always induces $n$-type conductivity with intrinsic and extrinsic
defects in spite of its amphoteric behavior as shown above. This
could explain the persistent $n$-type conductivity of ZnO at low
as well as high temperatures.
\par
In this letter we have shown that even the electrically-inactive molecular hydrogen is
shown to activate deep donor levels and thus bring in $n$-type
conductivity. H can exhibit cationic, anionic
and neutral character in a single material under different
conditions. The defect complexes formed
between H and prominent impurities like Li, B, and Al could
give rise to the H-related local vibrational modes in ZnO and
enhance its $n$-type conductivity. Therefore these impurities
should be removed by annealing treatments to bring in $p$-type
conductivity in ZnO. We have demonstrated that experimental
results together with present type of theoretical calculations are
needed to characterize impurities in semiconductors unambiguously.

The authors are grateful to the Research Council of Norway for
financial support (DESEMAT project under FRIENERGI program) and
computing time on the Norwegian supercomputer facilities (NOTUR). RV
thanks Institute of Mathematical Sciences for hospitality.


\begin{references}
\bibitem{chen05} Y.J. Chen, et.al.,
Appl. Phys. Lett. {\bf 87}, 233503 (2005).
\bibitem{beyer07}W. Beyer, J. Hupkes, and H. Stiebig, Thin Solid Films {\bf 516}, 147
(2007).
\bibitem{eizek08} J. \v{C}\'{i}\v{z}ek, et.al., J. Appl. Phys. {\bf 103}, 053508 (2008).
\bibitem{mollwo54} E. Mollwo, Z. Physik, {\bf 138} 478 (1954).
\bibitem{thomas56} D.G. Thomas and J.J. Lander, J. Chem. Phys.
{\bf 25}, 1136 (1956).
\bibitem{walle00} C.G. Van de Walle, Phys. Rev. Lett. {\bf 85}, 1012 (2000).
\bibitem{janotti07} A. Janotti and C.G. Van de Walle, Nature
Mater. {\bf 6}, 44 (2007).
\bibitem{shi04} G. A. Shi, M. Saboktakin, M. Stavola, and S.J.
Pearton, Appl. Phys. Lett. {\bf 85}, 5601 (2004).
\bibitem{wardle06} M.G. Wardle, J.P. Goss, and P.R. Briddon, Phys. Rev. Lett. {\bf 96}, 205504 (2006).
\bibitem{ip03} K. Ip, M. E. Overberg, et.al., Appl.
Phys. Lett. {\bf 82}, 385 (2003).
\bibitem{hutson57} A. R. Hutson, Phys. Rev. {\bf 108}, 222 (1957).
{\bibitem{hofmann02} D.M.Hofmann, A. Hofstaetter, et.al., Phys.
Rev. Lett. 88, 045504 (2002).
\bibitem{look03} D.C. Look, et.al., Physica B {\bf 340-342}, 32 (2003).
\bibitem{kassier08}  G. H. Kassier, et.al., Phys. Stat. Sol.
(c) {\bf 5}, 569 (2008).
\bibitem{lavrov09} E. V. Lavrov, F. Herklotz, and J. Weber, Phys. Rev.
Lett. {\bf 102}, 185502 (2009).
\bibitem{weber10} M.H. Weber, et.al., J. Electronic Mat. {\bf 39},
573 (2010).
\bibitem{herklotz10} F. Herklotz, E.V. Lavrov et.al. Phys. Rev. B
                   {bf 82}, 115206 (2010).
\bibitem{li08} X.-B. Li, S. Limpijumnong, et.al., Phys. Rev. B {\bf 78}, 113203 (2008).
\bibitem{lavrov05} E. V. Lavrov, F. B\"{o}rrnert, and J. Weber,
                   Phys. Rev. B {\bf 72}, 085212 (2005).
\bibitem{walle06} C.G. Van de Walle and J. Neugebauer, Annu. Rev.
                  Mater. Res. {\bf 36}, 179 (2006).
\bibitem{blochl94} P. E. Bl\"{o}chl, Phys. Rev. B {\bf 50}, 17953 (1994).
\bibitem{vasp}     G. Kresse and J. F\"{u}rthmuller, Comput. Mater. Sci. {\bf 6}, 15 (1996).
\bibitem{perdew96} J.P. Perdew, K. Burke, and M. Ernzerhof, Phys. Rev. Lett. {\bf 77},
                   3865 (1996).
\bibitem{heyd03} J. Heyd, G. E. Scuseria, and M. Ernzerhof, J. Chem. Phys. {\bf 118},
8207 (2003); {\bf 124}, 219906 (2006).
\bibitem{oba08} F. Oba, A. Togo, et.al., Phys. Rev. B {\bf 77}, 245202 (2008).
\bibitem{vidya11} R. Vidya, P. Ravindran, H. Fjellv{\aa}g, et.al., Phys.
Rev. B {\bf 83}, 045206 (2011).
\bibitem{ravi03} P. Ravindran, A. Kjekshus, H. Fjellv{\aa}g, et.al., Phys. Rev. B
67, 104507 (2003).
\bibitem{vaj05} P. Vajeeston, P. Ravindran, et.al., Euro. Phys. Lett. {\bf 72}, 569 (2005).
\bibitem{du11} M.-H. Du and K. Biswas, Phys. Rev. Lett. {\bf 106},
115502 (2011).
\bibitem{cox01} S.F.J. Cox, E.A. Davis, et.al., Phys. Rev. Lett.
{\bf 86}, 2601 (2001).
\bibitem{lavrov02} E. V. Lavrov, J. Weber, et.al., Phys. Rev. B {\bf 66}, 165205 (2002).
\bibitem{mccluskey02} M.D. McCluskey, et.al., Appl. Phys. Lett. {\bf 81}, 3807 (2002).
\bibitem{jokela07} S.J. Jokela and M.D. McCluskey, Phys. Rev. B {\bf 76}, 193201 (2007).

\bibitem{mccluskey07} M.D. McCluskey and S.J. Jokela, Physica B {\bf 401-402}, 355 (2007).
\bibitem{jokela03} S.J. Jokela, M.D. McCluskey, and K.G. Lynn, Physica B {\bf 340-342}, 221 (2003).
\bibitem{lavrov09_1} E.V. Lavrov, Physica B {\bf 404}, 5075 (2009).
\bibitem{lavrov03}  E.V. Lavrov, Physica B {\bf 340-342}, 195 (2003).
\bibitem{wardle05} M. G. Wardle, J. P. Goss, and P. R. Briddon, Phys. Rev. B {\bf 71}, 155205 (2005).
}
\end{references}
\end{document}